\documentclass[12pt,preprint]{aastex}

\newcommand{\mdot}{{\dot{\rm M}}}
\newcommand{\msun}{{~{\rm M}_\odot}}
\newcommand{\msunperyr}{{\msun {\rm yr}^{-1}}}

\revised{}
\accepted{}

\shorttitle{}

\shortauthors{Hicks et al.}

\received{2002 March 21}
\begin{document}

\title{Chandra X-Ray Spectroscopy and Imaging of the Galaxy Cluster
PKS 0745-191}

\author{Amalia K. Hicks, Michael W. Wise, John C. Houck and Claude R.Canizares}

\affil{MIT Center for Space Research, One Hampshire St., Building NE80-6007, Cambridge, MA 02139-4307}

\email{ahicks@alum.mit.edu, wise@space.mit.edu, houck@space.mit.edu,
crc@space.mit.edu}

\begin{abstract}

We present a detailed spectral and spatial analysis of the galaxy
cluster PKS 0745-191, using recent observations from the Chandra
X-ray Observatory.  The data provide information on the
temperature and metallicity of the intracluster medium, the
distribution of emission throughout the cluster, morphology of the
cluster core, and an independent mass estimate which can be
compared to that from gravitational lensing.  X-ray spectra
extracted from the central 300 kpc ($\sim$$2\arcmin$) are well
described by a two-temperature plasma and the mean cluster
temperature is consistent with previously determined values.  The
distribution of both temperature and metallicity within the inner
360 kpc ($2.3\arcmin$) of PKS 0745-191 is probed on scales of
$8\arcsec$ ($\sim$20 kpc), yielding a relatively constant
abundance throughout the region and a strong temperature gradient
(down to $\sim$4-5 keV) within the central 215 kpc.  $\beta$-model fits
to the surface brightness profile of PKS 0745-191 indicate that a
second $\beta$-model component is required to fit the inner
$\sim$10 kpc. Imaging analysis of the cluster core reveals an
irregular morphology within the inner 50 kpc, and confirms that
the X-ray emission peak lies within $0.5\arcsec$ (the spatial
resolution of Chandra) of both the optical and radio central cD
positions.  The cluster mass, based on both two temperature and
multiphase fits to the X-ray spectrum, is in close agreement with
lensing mass estimates of PKS-0745-191. 

\end{abstract}


\section{INTRODUCTION}
\label{s:intro}

As the largest gravitationally bound objects in the universe,
clusters of galaxies are important tracers of the large scale
distribution of matter.  Because the distribution of clusters
depends sensitively on the underlying cosmology
\citep{ev90,frenk90,rlt92,nfw95,ecf96}, improved tests of
cosmological models require accurate measurements of the
properties of many individual clusters. The masses of galaxy
clusters and their internal matter distribution are of particular
interest.  Three primary observational techniques have been widely
used to derive the mass distribution in galaxy clusters:
measurements of the galaxy velocity dispersion, modeling of
gravitational lensing effects, and examination of X-ray emission
from hot intra-cluster gas.

Most early measurements of cluster masses were based on optical
observations of the motion of cluster galaxies. These measurements
are complicated by the fact that cluster membership for individual
galaxies is often uncertain. Furthermore, although determination
of cluster masses from the galaxy velocity dispersion is founded
on the assumption that the cluster is virialized, the
virialization requirement is often unsatisfied. Many clusters are
dynamically young and unrelaxed because the dynamical time scale
for cluster formation is comparable to the Hubble time
(Richstone et al. 1992). 

Gravitational lensing studies infer the cluster mass from
measurements of distortions produced in images of background
galaxies.  Lensing studies have the advantage that they directly
measure the total projected mass in all forms along the line of
sight. In particular, the lensing mass includes both luminous and
dark matter, independent of assumptions about the dynamical state
of the cluster \citep{tvw90,ks93}.  However, lensing measurements
give no information on the dynamical state of the cluster or the
relative amounts of luminous and dark matter or the segregation
of baryonic matter between galaxies and intra-cluster hot gas.

X-ray observations of clusters provide complementary data,
yielding a measure of the cluster mass which is independent of
mass concentrations along the line of sight and also, through the
plasma temperature map, yielding information on the dynamical
state of the cluster \citep{me01}.  The intracluster gas provides a useful
probe of the cluster mass because the gas rapidly attains
hydrostatic equilibrium in the cluster potential.  In equilibrium,
the gas temperature is typically in the range $10^7 - 10^8$ keV,
making the gaseous intracluster medium a prodigious source of
thermal X-ray emission. Using X-ray observations to measure the
gas density and temperature and assuming spherical symmetry, the
radial mass distribution, $M(r)$, can be derived from the
condition of hydrostatic equilibrium
\begin{equation}
M_{\rm tot}(<r) = - {kT r \over \mu m_p G}
    \left [
    {{\rm d\,ln}~\rho_{\rm gas} \over {\rm d\,ln}~r} +
    {{\rm d\,ln}~T \over {\rm d\,ln}~r} 
    \right ],
\label{hydrostat}    
\end{equation}
where $\rho_{\rm gas}$ is the gas density and $T$ is the gas
temperature. Detailed simulations indicate that the assumption of
hydrostatic equilibrium can yield reasonably accurate cluster
masses even in cases where the cluster has not completely relaxed,
although the exact range of validity of this approach remains
uncertain \citep{sch96,emn96,an96,enf98}.

Along with their cosmological implications, X-ray observations of
galaxy clusters are also motivated by interest in the origin and
evolution of the gaseous intracluster medium itself.  One question
which has been the focus of intensive effort for a number of years
concerns the fate of the X-ray emitting gas.  In a large fraction
of dynamically relaxed clusters, the radiative cooling time of gas
in the cluster core is short compared to the Hubble time.  In the
absence of a heat source capable of balancing losses due to
cooling, the intracluster gas is expected to settle inward in a
``cooling flow'' (e.g. see \citet{cls88, fab94}).  Previous X-ray
estimates of the cooling rates ($\mdot$) revealed many clusters
with $\mdot > 100 \msunperyr$ and, in some cases, as high as
$\mdot \sim 1000 \msunperyr$. Once it cools below X-ray emitting
temperatures, the fate of the gas is unclear but, given the amount
of material involved, some observable signature is expected.
Sustained over a Hubble time, these cooling rates imply the
accumulation of $> 10^{12} \msun$ of material, but detailed
examination of clusters at many wavelengths has so far failed to
confirm the amount of cooling material inferred from X-ray
observations or to reveal its final state.

\citet{allen-PKS} and \citet{allen} have argued that
the presence of a cooling flow can have an important affect on the
cluster mass determination. Recent comparisons between cluster
masses determined from X-ray observations and from gravitational
lensing have revealed disagreements in the sense that the lensing
value is sometimes larger than the X-ray determined value.
However, Allen et al. (1996) have argued that, at least for relaxed
clusters, better agreement can be obtained if the presence of a
cooling flow is taken into account.

The luminous cluster PKS 0745-191 is an excellent example.  Using
X-ray observations from ASCA and ROSAT, Allen et al. (1996) report
that PKS 0745-191 contains one of the largest known cooling flows,
with a mass deposition rate of $\sim 1000 \msunperyr$.  From
optical observations, they also report the discovery of a bright
arc consistent with gravitational lensing of a galaxy at redshift
$z=0.433$.  Using the lensing data, they derive the projected mass
of the central part of the cluster, finding good agreement with
the corresponding mass obtained from a multiphase analysis of the
X-ray data.  They emphasize, however, that a multiphase analysis
is required; their single-phase analysis of the ASCA data clearly
disagrees with their lensing result, yielding a mass which is
smaller than the lensing value by a factor of $\sim 3$.

In this paper, we present our analysis of high-angular resolution
X-ray imaging observations of PKS 0745-191 using the Chandra X-ray
observatory.  From these data, we derive mass estimates and
cooling flow mass deposition rates.  X-ray surface brightness and
abundance profiles are also presented, along with a map of the
projected plasma temperature distribution. In addition, we compare
our high-resolution X-ray image of the cluster core with
observations in the optical and radio.

Throughout this paper we used $\rm{H}_0$=50\hspace{.05in}
$\rm{km\hspace{.05in}s^{-1}\hspace{.05in}Mpc^{-1}}$, 
$\Omega_M$=1.0, $\Omega_\Lambda$=0.0, and $\rm{q}_0$=0.5.  For PKS
0745-191 we take the redshift to be $z=0.1028$, so that the
luminosity distance $D = 631$ Mpc, and $1\arcsec$  corresponds to
a linear distance of 2.57 kpc.  

\section{OBSERVATIONS}
\label{s:obs}

PKS 0745-191 was observed twice using the Chandra/HETGS.  The
first of these observations, observation ID (obsid) 510, occurred
on October 14, 1999, when the temperature of the ACIS focal plane
was $-110^o$ C.  The second observation (obsid 1509) was
on March 4, 2000 at a focal plane temperature of -120.  There
was a 208 degree difference in roll angle between the two
observations.  

The ACIS gain was corrected using the most recent calibration
files available, taking into account the focal plane temperature
of the instrument during each observation.  The aspect solutions
were examined for irregularities but none were found.  Background
contamination due to charged particle flares was reduced by
removing time intervals during which the background rate exceeded
the average background rate by more than 20\%.  The event files
were filtered on standard grades and bad pixels were removed using
standard cleaning procedures. After filtering, the net remaining
exposure times were 44784 seconds and 38999 seconds, respectively.

We fit a two-dimensional elliptical Lorentzian to the counts image
of each dataset to locate the center of the X-ray emission peak
and measure its ellipticity. The center positions of the two
observations differ in declination by $3\arcsec$, apparently due
to an astrometry error in pipeline processing.  For this reason,
our imaging analysis used only the more recently reprocessed
observation, obsid 510.  Our spectral analysis is unaffected by
the offset.

Using obsid 510, the centroid of the flux corrected X-ray emission
lies at (07:47:31.3$\pm{0.5}$, -19:17:40.0$\pm{0.5}$) (J2000), the
exact coordinates of the cD galaxy.   
The axis-ratio is $0.65\pm{0.06}$, with the major axis at position angle
of $99^\circ\pm{3}$, measured counterclockwise on the sky from North.
      
The zeroth order image is sufficiently extended that it overlaps
with events from the dispersed spectra. This overlap has no effect
on the cluster core, becoming signifcant only at a radius of
$75\arcsec$ from the center of the zeroth order image. The
cross-dispersion extent of the overlap region was measured by
fitting Gaussians to the cross-dispersion profiles.  To remove the
overlap, we excised the region beginning $75\arcsec$ (193 kpc)
along the dispersion direction from the cluster center, and
extending to approximately 2$\sigma$ in the cross-dispersion
direction on either side of the axis of the dispersed spectrum.
Within a radius of $2.3 \arcmin$ (360 kpc) any remaining contamination
lies at a very low level and is restricted to  energies $E > 4$ keV
along the MEG dispersion direction and $E > 8$ keV along the HEG
dispersion direction. 

\section{SPECTRAL ANALYSIS}
\label{s:spec-analy}

Chandra's unparalleled spatial resolution makes it possible to 
examine spatial variations in the X-ray emission on much smaller
scales than were previously accessible. In this section we present
our analysis of the structure of the inner $2.3\arcmin$ (360 kpc)
of PKS 0745-191. We also discuss a similar analysis of the
emission-weighted mean spectrum which may be directly compared
with results from lower resolution observations.

\subsection{Integrated Spectrum Analysis}
\label{ss:spec-analy}

Spectra were extracted from both data sets in an elliptical region
with a $2\arcmin$ ($\sim$300 kpc) major axis and an ellipticity of
0.35, centered on the X-ray emission peak. These extractions yielded
spectra with 56,863 counts and 47,288 counts for OBSID 510 and 1509,
respectively. These spectra were then analyzed with XSPEC
\citep{xspec} using response matrice files (RMFs) and effective area
files (ARFs) generated from the latest version of the Chandra
calibration database (CALDB 2.7);  the RMF includes the improved
calibration for the low energy response of the aimpoint CCD on
ACIS-S. For all analyses, a single RMF was generated at the position
of the X-ray centroid and used for analysis of the entire region.
Variations in the spectral resolution of the ACIS-S aimpoint CCD were
found to be smaller than 10\% within the extraction region; fits using
RMFs taken from other points in the region gave equivalent results
within the errors. Variations in the effective area were
negligible within the extraction region. Backgrounds were
extracted for these observations from the aim point chip
(ACIS-S3).  The background regions were chosen as far away from
the aimpoint as possible and were sized to minimize contamination by 
diffracted photons. Each spectrum was grouped to
contain least 20 counts per bin, and data from the two
observations were fit simultaneously.  

The spectra were fitted with three models, each including foreground
absorption: a single temperature model, a two temperature model, and a 
single temperature plus cooling flow model.  Each of the spectral
models were fit twice, once with the absorbing column frozen at its
measured value of 4.24 $\times{10}^{21}\rm{{cm}^{-2}}$ \citep{nh},
and once allowing the absorption to vary.  
The cluster redshift was fixed throughout the analysis.  
Data below 0.5 keV were excluded from the fits due to uncertainties
in the ACIS calibration while energies above 8.0 keV were neglected
due to background contamination.  The results of these spectral fits
along with 90\% confidence ranges, are shown in {Table-1}.   

Of the models considered, the best fits were obtained with two
temperature MEKAL and multiphase MKCFLOW plasma models.  A plot of
the fit to the two temperature model is shown in {Figure-1}.  For this 
model, we find an ambient cluster temperature of ${8.7}^{+0.5}_{-0.5}$
keV, a low temperature component of ${1.2}^{+0.1}_{-0.1}$ keV, and an
abundance of ${0.52}^{+0.05}_{-0.05}$ solar. The reduced $\chi^2$ for
this fit  was 1.02 for 929 degrees of freedom.  These results are
consistent with both the $0-2\arcmin$ ASCA cooling flow analysis of
Allen et al. (1996) and the $0-2\arcmin$ BeppoSAX multiphase analysis
of PKS 0745-191 \citep{DM99}.     
 
The larger residuals between 1.3-2 keV {Figure-1} appear to be the
result of a calibration problem, most likely associated with spatial
variations in the RMF combined with interpolation errors on the rather
coarse energy grid used in the RMF calibration. Similar residuals are
seen in observations of other clusters, generally appearing in cases
where spectra have been extracted from a source which spans several of
the detector regions where calibration data is specified. These
residuals do not significantly affect our fit results.

\subsection{Spatially Resolved Spectra}
\label{ss:resolved-spec}

To investigate the radial dependence of temperature and abundance
in PKS 0745-191, spectra were extracted from 12 elliptical annuli
spanning the central $2\farcm 3$ (360 kpc) of the cluster.
{Figure-2} shows the spectral extraction regions overlaid on an 
adaptively smoothed counts image of the cluster.  The extraction
regions were sized to ensure at least 3000 counts per spectrum in
the 0.29-7.0 keV band for each data set.  The extracted spectra
were grouped to contain 20 counts per bin and analyzed using
XSPEC. Models were fit to data over the energy range 0.6-7.0 keV.
ARFs, RMFs, and backgrounds used were the same as described in
\S\ref{ss:spec-analy}, and data from the two observations were again fit
simultaneously at each radius.  

The same plasma models described in \S\ref{ss:spec-analy} were again
used to fit the data.  While single temperature spectral models
generally fit the data well (reduced $\chi^2$ of 0.98-1.55),
adding a cooler component in fits to the nine innermost
spectra (the central 215 kpc of the cluster) improved the fits
somewhat.
Outside of that region, the normalization of the low
temperature component of the two temperature model was consistent
with zero, and single temperature models provided equally good
fits to the data.
The cooling flow model did not constrain the temperature outside of
the innermost region.  

Because of large discrepancies in spatial resolution, it is
difficult to compare prior work on the spatial structure of PKS
0745-191 to the results reported in this section.  However, it is
worth noting that our results differ in some respects.

In their BeppoSAX analysis, \citet{DM99} ruled out a large
temperature gradient in PKS 0745-191, but our investigation shows
that there is indeed a noticeable gradient present in the central
region (Figures 3 \& 4).  Their work also supports a strong
abundance gradient, but we see no indication of such a gradient in
the inner 360 kpc ($2.3\arcmin$) of the cluster.  Instead, we find
that the abundance remains relatively constant throughout the core
({Figure-5}), with an average value of $\sim 0.5$ solar,
consistent with their result for the inner $2\arcmin$.  The
results of Allen et al. (1996) are also consistent with a constant
metal abundance although they favor a somewhat lower value of
$\sim 0.3$.

\subsection{Temperature Maps}

Extending the radial analysis discussed above, we have generated
maps of the plasma temperature in the central regions of the
cluster. These maps were computed for both datasets independently 
using a custom module within the ISIS \citep{houck} spectral
analysis package which automates the process of extracting many 
PI spectra, generating appropriate ARFs and RMFs for each
spectrum, and then fitting a given spectral model (see
\citet{tmap} for a detailed description).
 
In generating the temperature maps, a grid of adaptively sized
extraction regions were selected to contain a minimum of 2000
counts in the 0.3-7.0 keV band. The grid was selected to span a
200x200 arcsec region centered on the cluster peak, consistent
with the integrated spectrum analysis described above. We avoid
problems associated with chip edge effects by choosing spectral
extraction regions which lie entirely within the edges of the
ACIS-S3 chip.  The extracted spectra were fit with a MEKAL plasma
model including foreground Galactic absorption fixed at the
nominal value of 4.24 $\times{10}^{21}\rm{{cm}^{-2}}$.  The
abundance was fixed at a value of 0.5 as determined by the
integrated spectral fits.  Single-parameter 90\% confidence limits
were computed for each fit to estimate the level of statistical
uncertainty in each pixel of the temperature map. Although the
temperature map shown in {Figure-6} was computed holding the
absorbing column and metal abundance fixed, similar results were
obtained when the absorbing column and metal abundance were
allowed to vary.

The temperature map for the central 100$^{''}$x100$^{''}$ region
of PKS 0745-191 (OBSID 510) is shown in {Figure-6}.  In general,
the resulting temperature structure is consistent with the radial
temperature profile presented in {Figure-3}. The overall
temperature morphology traces the X-ray surface brightness.  The
region of coolest emission, with temperatures $\sim$3-4 keV, appears
to be centered at a slight offset from the position of the central cD.
However, this scale is comparable to the map resolution in the core
and so may be consistent with a zero offset. Finally, we find no
evidence for gas cooler than $\sim$3 keV consistent with results from
other cooling flows observed with Chandra and XMM \citep{peterson,wise}. 

\section{SURFACE BRIGHTNESS}
\label{s:surf-br}

Radial surface brightness profiles were computed for each observation. 
To minimize errors due to the energy-dependence of the effective
area, we divided the 0.29-7.0 keV bandpass into 12 bands in which
the effective area is roughly constant.  In each energy band,
we extracted a 2048$\times$2048 image of the undispersed zero-order counts.
We also computed a zero-order exposure map for each energy band.
After excising regions affected by contamination from 
dispersed counts, the surface brightness was computed in
elliptical annuli with centroid, position angle and axis-ratio
given in \S \ref{s:obs}.  All quoted radii are measured along the
semi-major axis.

The surface brightness profiles from each dataset were mutually
consistent within their errors.  In each energy band, the surface
brightness profiles from each dataset were fit simultaneously,
once with a single $\beta$-model and once with a double
$\beta$-model. The single $\beta$-model has the form 
\begin{equation}
I(r) = I_B + I_0 \left( 1 + {r^2 \over r_0^2} \right)^{-3\beta+\frac{1}{2}}
\label{sb_eq}
\end{equation} 
where $I_B$ is a constant representing the intensity
contribution of the background, 
$I_{0}$ is the normalization and $r_{0}$ is the core
radius.
The double $\beta$-model has the form
\begin{equation}
I(r) = I_B +
       I_{1} \left( 1 + {r^2 \over r_{1}^2} \right)^{-3\beta_1+\frac{1}{2}} + 
       I_{2} \left( 1 + {r^2 \over r_{2}^2} \right)^{-3\beta_2+\frac{1}{2}}
\label{sbdoub_eq}
\end{equation}
where each component has fit parameters $(I_k, r_k, \beta_k)$.

{Table-2} shows the results of fitting a single $\beta$-model to
those energy bands which have sufficient signal to noise to
constrain the parameters.  Although a single $\beta$-model yields
a reasonably good fit to the surface brightness profile, it 
fails to account for a
significant excess of emission in the cluster core.  Adding a
second $\beta$ component ({Table-3}) yields a better fit
to the surface brightness profile ({Figure-7}), consistent with
the findings of \citet{cent} and \citet{doubeta}.  

The double $\beta$-model yields the best fit to the 0.29-7.0 keV
data giving $\beta = 0.49^{+0.02}_{-0.01}$ (at large radius), a core
radius of $51.42^{+2.57}_{-2.57}$ kpc, and a background value of
1.4$\times10^{-8}$  phot  $
~\rm{s}^{-1}$ $\rm{cm}^{-2}$ $\rm{arcsec}^{-2}$, with
a reduced $\chi^2$ of 1.15 for 473 degrees of freedom.   
Although the core radius of our best fitting single
$\beta$-model is comparable to the $37.5\pm{5}$ kpc core radius
reported by \citet{allen}, the outer core radius of the double
$\beta$-model is significantly larger.  

\section{IMAGING ANALYSIS}
\label{s:imag}  

\subsection{Substructure}
\label{ss:imag-substr}

We used an unsharp masking technique to examine the substructure
in the cluster core. By subtracting a heavily smoothed,
flux-corrected image ($I_S$) from an adaptively smoothed, full-resolution
image ($I_F$), we effectively subtract out the smooth, large scale
structure leaving residuals associated with smaller-scale
inhomogeneities.

A full resolution 400$\times$400 pixel image centered on the X-ray
emission peak was extracted from obsid 510 in the 0.29-7.0 keV
band. A matching zero-order exposure map was constructed for the
same region. Regions contaminated by dispersed photons were
removed from both the image and exposure map and the resulting
image was adaptively smoothed using the CIAO tool CSMOOTH.  A
similar image was smoothed with a Gaussian kernel of 20 pixel
($10\arcsec$) FWHM.  Each of the two resulting images was divided
by the exposure map and exposure time, to produce flux-corrected
images. The residual image, ($I_F - I_S$), with hardness contour
overlays is shown in {Figure-8}. 
     
Substructure is clearly present in the core of PKS 0745, with
excess emission covering an elongated area of about 100 kpc
$\times$ 50 kpc.  The brightest point in this image lies directly
on the cD galaxy at the center of the cluster, and has a surface
brightness which is 68\% of the adaptively smoothed surface
brightness at that point.  

Similar images were made in a hard band ($H$: 2.0-7.0 keV) and a soft
band ($S$: 0.29-2.0 keV).  These images showed no appreciable
difference in the morphology of the core substructure, although
the relatively low signal to noise ratio is not sensitive to weak
temperature variations. 

\subsection{Hardness Map}
\label{ss:hardness}
Images with the same specifications as those mentioned in \S
\ref{ss:imag-substr} were made of both the hard band ($H$) and the
soft band ($S$).  These images were smoothed using the smoothing
scale produced in the analysis of the full energy band (0.29-7.0
keV).  Using exposure maps calculated for the individual bands,
and following the procedure detailed in \S\ref{ss:imag-substr},
smoothed, flux-corrected images were constructed.  Using these
flux-corrected images $I_H$ and $I_S$, we constructed a
hardness map (${\cal H}$), defined as
\begin{equation}
  {\cal H} \equiv {I_S - I_H \over I_S + I_H}.
\end{equation} 
Contours of the resulting
hardness ratio image (${\cal H}$) are overlayed on the core
residuals map of {Figure-8}. 

Although the hardness contours are less elongated than the
residuals image, ellipticity is clearly present.  Soft emission is
peaked in two places, neither of which corresponds exactly to the
position of the cD, although the more central of the soft emission
peaks lies within $3\arcsec$ of it. 

\subsection{Optical and Radio Comparisons}

The adaptively smoothed core image ($I_F$) from
\S\ref{ss:imag-substr} was used to construct a set of ten X-ray
isophotes, spaced logarithmically over a range in surface
brightness from 2.2$\times10^{-8}$ to 1.8$\times10^{-6}
\hspace{0.05in} \rm{phot\hspace{0.03in} s^{-1}
\hspace{0.03in}cm^{-2} \hspace{0.03in}pix^{-2}}$.  {Figure-9}
shows these contours overlaid on an HST optical image of PKS
0745-191 obtained from the HST archive.
  
Ten isophotes were also constructed with the 8350 Mhz VLA radio image
of PKS 0745-191 \citep{radio}, logarithmically spaced over the range
1$\times10^{-4}$ to 1$\times10^{-2} \rm{Jy \hspace{0.05in}
beam^{-1}}$.  {Figure-10} shows these radio contours overlaid on the
adaptively smoothed core image ($I_F$) from \S\ref{ss:imag-substr}.  

Figures 9 and 10 show that the X-ray emission peak of PKS 0745-191
lies within $0.5\arcsec$ (the spatial resolution of Chandra) of both
the optical and radio centers of the cD galaxy.  This is
consistent with the ROSAT result of Allen et al. (1996) who
found that their X-ray emission peak corresponded to the optical
position of the cD within the point spread function of ROSAT
($\sim$$5\arcsec$).   

Examination of the center of the optical image, reveals that the
major axis of the cD is misaligned with the X-ray isophotes,
although the optical and X-ray contours are well correlated on
larger scales. In contrast, the radio morphology does not appear
correlated with the X-ray emission. An anti-correlation between
X-ray and radio morphology has been seen in some nearby clusters
of galaxies \citep{mcnamara}, but these data do not clearly show
this effect.

The brightest of the gravitiational arcs of PKS 0745-191 can be seen
in the optical image as a faint streak directly to the right of
the cluster.  There seems to be no apparent connection between the
X-ray emission and the arcs. 

\section{MASS DETERMINATION}
\label{s:mass}

\subsection{Density and Cooling Time}

A cluster of galaxies whose X-ray surface brightness profile 
follows a $\beta$-model is consistent with a gas density
distribution of the form
\begin{equation}
 n_{\rm gas}(r) = n_0 \left(1 + {r^2 \over r_c^2}\right)^{-3\beta/2}
\label{dens}
\end{equation}
where $n_0$ is the cluster central density.  
This relation assumes isothermal, hydrostatic equilibrium in
spherical symmetry.  X-ray spectral fits determine both plasma
temperature and, via the emission measure, the central density.  

Spectra were extracted from an elliptical aperture with a 0.5 Mpc
major axis, which was centered on the X-ray emission peak.  The
spectra were grouped into 20 count bins, and the 0.5-8.0 keV data
were fit in XPEC with the same calibration files, backgrounds, and
models employed in \S \ref{ss:spec-analy}.  The resulting best-fit
temperatures (using the higher temperature from the two temperature
model) and computed central densities are given in {Table-4}.    
     
The characteristic time that it takes a plasma to cool isobarically
through an increment of temperature $\delta$T can be written
\begin{equation}
  \delta t_{cool} = \frac{5}{2}~\frac{k}{n~\Lambda(T)}~\delta T 
\label{cool}
\end{equation}
where $n$ is the electron density, $\Lambda$(T) is the
total emissivity of the plasma (the cooling function), and k is
Boltzmann's constant.  For isochoric cooling, 5/2 is
replaced by 3/2. 

Using this expression, the distribution of cooling times can be 
derived from the cluster density profile.  The radius at which the
cluster cooling time is equal to the cluster age (generally
assumed to be a large fraction of the Hubble time) is defined as
the cooling radius.  Using $H_0=$50, $q_0=$0.5, the assumption of
isobaric cooling, our $\beta$-model fit parameters, and the fit
results of the cooling flow model, we determined the cooling radius
and central cooling time of PKS 0745-191.  For a cluster age of
1$\times10^{10}$ yrs, and using the cooling function appropriate to a
0.5 solar abundance plasma, we get a cooling radius of 104 kpc and a
central cooling time of 8.67$\times10^8$ years in the cluster core.   

Allen et al. (1996) calculate a cooling radius of $180^{+11}_{-9}$
kpc from a ROSAT HRI observation, and $212^{+52}_{-23}$ kpc from
ROSAT PSPC data.  Using a deprojection analysis, they derive
somewhat longer cooling times of 1.95$\times{10}^9$ and
1.11$\times{10}^9$ years, respectively.  However, their analysis
assumed an ambient cluster temperature of 10 keV, and the size of
the central bins from which they extract cooling times are
relatively large: $15\arcsec$ for the PSPC, and $8\arcsec$ for the
HRI.  Given that our results are roughly consistent at a radius of
200 kpc, the detailed differences probably arise from a
combination of the $\sim$2 keV temperature difference and the
difference in the bin size.  

\subsection{Mass Comparisons}

Optical HST images indicate the presence of a gravitational
lensing arc at a radius of 45.9 kpc from the center of PKS
0745-191.  This arc, likely due to lensing of an
early-type spiral galaxy at a redshift of 0.433, has a length of
$\sim$$15.5\arcsec$ and width of $\sim$3.5 arcsec.  From a
detailed lensing analysis Allen et al. (1996) derived a spherically
projected mass of 3.0$\times10^{13}~M_{\odot}$, and an elliptical mass
estimate of 2.1-2.5$\times10^{13}~M_{\odot}$ within the tangential
critical radius.   

Using our $\beta$-model fit parameters, the spectral fitting results
of {Table-4}, equation 2, and the equation of hydrostatic
equilibrium (\ref{hydrostat}) 
we have computed various mass estimates for PKS 0745-191.   
{Table-4} gives the gas mass within 0.5 Mpc
of the core, the total mass within 0.5 Mpc, and the lensing mass.
Lensing mass estimates were calculated for a cylindrical region
with radius 45.9 kpc and an assumed cluster extent of 1 Mpc.
Extending the cluster further does not significantly affect
the results.   

The X-ray mass estimate based on a multiphase analysis is
consistent with spherical optical mass estimates, while the two
temperature model provides a mass comparable to elliptical lensing
estimates.  In contrast, the single temperature model clearly
underestimates the mass.  Our cooling flow derived lensing mass
estimate is consistent with that reported by \citet{allen}.

\subsection{Mass Estimates}

Although it provides an excellent description of the X-ray
surface-brightness profile (see Fig 6), the hydrostatic,
isothermal, spherically symmetric $\beta$-model is an obvious
oversimplification, given the observed temperature gradient and
elliptical morphology. Nevertheless, the mass estimates are
relatively insensitive to these details.

\citet{ev90} investigated the accuracy of
mass estimates based on the $\beta$-model using a large number of
cosmological gas-dynamic simulations of cluster formation and
evolution. He concluded that $\beta$-model mass estimates are
reasonably accurate when computed for regions with mean density
($\bar{\rho}$) between 500 and 2500 times the critical density
($\rho_c = 3 {H_0}^2 / 8 \pi G$). Using our mass estimates (see Table
5), the region inside a 0.5 Mpc radius has a mean density of
$\bar{\rho} \gtrsim 1300 h^{-2} \rho_c$, placing it within the
hydrostatic region commonly seen in cluster evolution simulations.

Using his simulations, \citet{ev90} also found that the presence of
a temperature gradient has a relatively small effect on the mass
determination.  After including the known 3D temperature gradient
from the simulated clusters, the corresponding mass estimates
for regions with $\bar{\rho} \gtrsim 500$ changed by less than 1
standard deviation.  This is consistent with the conclusion
obtained by comparing the sizes of the temperature and density
gradients in equation (\ref{hydrostat}). Both the
single-temperature fits and the hotter component of the
two-temperature fits give a temperature gradient of about
${\rm d\,ln\,}T/{\rm d\,ln\,}r \approx 0.3$ at a radius of 200 kpc.  For
comparison, the $\beta$-model indicates a gas-density gradient of
${\rm d\,ln\,}\rho_g/{\rm d\,ln\,}r \approx 1.3$ at the same radius.
Therefore, in estimating the total mass within this radius, the
temperature gradient is at most a 25\% effect.

A somewhat larger source of uncertainty arises from the cluster's
elliptical shape. Although the morphology may be more complex, we
can estimate the uncertainty due to projection effects by
computing the gas mass for oblate and prolate mass distributions
suitably oriented and normalized so that the surface brightness
and axis-ratio matches the observations. For an axis-ratio of 0.6,
the estimated gas mass could vary by as much as $\sim$50\%
depending on the geometry, although the extremes of this range
require coincidental alignment, making them less likely.

\section{SUMMARY AND DISCUSSION}
\label{s:summary}

We have presented our analysis of Chandra observations of the
galaxy cluster PKS 0745-191.  These data provide a much clearer
picture of the X-ray structure of the central 300 kpc.  The spectra
are not well fit by a single temperature model, because 
of excess soft emission from the core.  The
cluster-integrated spectrum is best fit by a two-temperature model
with temperatures of ${8.7}^{+0.5}_{-0.5}$ keV and
${1.2}^{+0.1}_{-0.1}$ keV, with an abundance of
${0.52}^{+0.05}_{-0.05}$ solar. Consistent with results
from other Chandra observations of cooling flows, we see no 
evidence for X-ray emission from cooling gas with temperatures
below $\sim 2-3$ keV \citep{peterson,wise}.

Chandra's excellent angular resolution has allowed us to study the
distribution of both metallicity and temperature on very small scales
($\sim$20 kpc) in the inner 360 kpc of PKS 0745-191.  Using single
temperature spectral models we find an ostensibly constant abundance
with a value of $\sim$0.5 solar, and a steep temperature gradient
within the innermost 215 kpc.

Imaging analysis of the cluster core reveals enhanced X-ray
emission from an elongated region in the core of about 100 kpc
$\times$ 50 kpc. The hardness ratio map shows a somewhat larger
region (R $>$ 50 kpc) of excess soft emission (0.29-2.0 keV) in
the core. Comparison of the X-ray data to optical (HST) and radio
(VLA) observations of PKS 0745-191 places the X-ray emission peak
within $0.5\arcsec$ of both the optical and radio central cD
positions.  

The surface brightness profile is best fit by a double $\beta$-model
with an outer core radius of 51.42 kpc and $\beta$ value of
0.49.  The central density is $\rm{n_0} = 0.069_{-0.001}^{+0.001}$.  The
central cooling time based on a multi-phase cooling flow model is
8.67$\times10^8$ yr, with a corresponding cooling radius of 104
kpc.  

We found best agreement between the X-ray and spherical gravitational
lensing masses using a multiphase spectral model for which we
obtained a mass of 2.9$\times 10^{13}~\rm{M}_{\odot}$.  Single
temperature models imply masses consistently smaller than lensing mass
estimates.  Mass results stemming from two temperature models, though
smaller than spherical lensing estimates, are consistent with
elliptically derived values.  In any case, our X-ray spectra require
no more than two temperature components and, in particular, do not
require a multiphase treatment of the sort emphasized by
Allen et al. (1996).  Although the physics of cooling gas strongly
suggests that a multiphase medium should be present, our data do not
require it.  

\acknowledgments

We thank Greg Taylor for providing access to his radio observations of PKS
0745-191.

\clearpage

\begin{deluxetable}{ccccccccc}
\tablecaption{Comparison of Integrated Spectral Fits for R $<$ 300 kpc \label{Table-1}}
\tablewidth{0pt}
\tablehead{
\multicolumn{2}{c}{Model} & \colhead{kT} & \colhead{$\rm{N_H}$} & \colhead{Z} & \colhead{$\dot{\rm{M}}$} & \colhead{$\rm{kT_{low}}$} & \colhead{$\chi^2$/\rm{DOF}} \\
\multicolumn{2}{c}{} & \colhead{[keV]} &
\colhead{[${10}^{21}\rm{{cm}^{-2}}$]} &\colhead{[solar]} &
\colhead{[$\rm{M}_{\odot} \rm{{yr}^{-1}}$]} & \colhead{[keV]} & \colhead{} \\
} 
\startdata
\multicolumn{2}{c}{Single T} & {${6.9}^{+0.2}_{-0.2}$} & 4.24 & {${0.43}^{+0.04}_{-0.04}$} & \nodata & \nodata & 1101/931 \\
\vspace{.001in} \\
\multicolumn{2}{c}{} & {${8.1}^{+0.4}_{-0.4}$} & {${3.4}^{+0.1}_{-0.1}$} & {${0.49}^{+0.05}_{-0.05}$} & \nodata & \nodata & 988/930 \\
\vspace{.1in} \\
\multicolumn{2}{c}{2T} & {${8.7}^{+0.5}_{-0.5}$} & 4.24 & {${0.52}^{+0.05}_{-0.05}$} & \nodata & {${1.2}^{+0.1}_{-0.1}$} & 946/929 \\
\vspace{.001in} \\
\multicolumn{2}{c}{} & {${8.8}^{+0.5}_{-0.5}$} & {${4.1}^{+0.3}_{-0.3}$} & {${0.52}^{+0.05}_{-0.05}$} & \nodata & {${1.3}^{+0.1}_{-0.2}$} & 945/928\\
\vspace{.1in} \\
\multicolumn{2}{c}{1T + CF} & {${12.1}^{{+3.00}}_{-0.7}$} & 4.24 & {${0.55}^{+0.06}_{-0.06}$} & {${970}^{+67}_{-59}$} & 0.001 & 947/930 \\
\vspace{.001in} \\
\multicolumn{2}{c}{} & {${12.1}^{{+3.6}}_{-0.9}$} & {${4.27}^{+0.09}_{-0.09}$} & {${0.55}^{+0.06}_{-0.06}$} & {${985}^{+72}_{-55}$} & 0.001 & 947/929 \\
\enddata
\end{deluxetable}

\clearpage

\begin{deluxetable}{crrrrr}
\tablecolumns{6}
\tablewidth{0pc}
\tablecaption{Beta-Model Fits\label{Table-2}}
\tablehead{ 
\colhead{E [keV]} & \colhead{$R_{\rm core}$} [\arcsec] & \colhead{$\beta$}  & \colhead{$I_{\rm 0}$\tablenotemark{a}} & \colhead{$I_{\rm B}$\tablenotemark{a}}  & \colhead{$\chi^2$/DOF}             } 
\startdata 
0.29--7.00 & $18.0_{-0.6}^{+0.4}$ & $0.47_{-0.01}^{+0.01}$ &
$3919_{-88}^{+104}$ & $11_{-2}^{+2}$ & 616.3/476  \\ 
\vspace{.001in} \\
0.29--2.00 & $15.2_{-0.9}^{+0.9}$ & $0.48_{-0.01}^{+0.01}$ & $2677_{-120}^{+126}$ & $8_{-1}^{+1}$ & 557.8/476  \\ 
\vspace{.001in} \\
2.00--7.00 & $19.6_{-0.9}^{+0.9}$ & $0.47_{-0.01}^{+0.01}$ & $1761_{-54}^{+57}$ & $5.6_{-0.9}^{+1.3}$ & 567.3/476  \\ 
\vspace{.025in}\\
\cline{1-6}\\
0.54--1.00 & $13_{-1}^{+1}$ & $0.48_{-0.02}^{+0.01}$ & $828_{-80}^{+77}$ & $3.6_{-0.6}^{+1.5}$ & 427.5/476  \\ 
\vspace{.001in} \\
1.00--1.57 & $16_{-1}^{+1}$ & $0.49_{-0.01}^{+0.01}$ & $1054_{-68}^{+71}$ & $1.3_{-0.2}^{+0.9}$ & 532.4/476  \\ 
\vspace{.001in} \\
1.57--2.00 & $17_{-1}^{+2}$ & $0.48_{-0.01}^{+0.02}$ & $553_{-38}^{+41}$ & $1.5_{-0.2}^{+0.6}$ & 504.2/476  \\ 
\vspace{.001in} \\
2.00--3.50 & $19_{-1}^{+1}$ & $0.48_{-0.01}^{+0.01}$ & $1009_{-44}^{+47}$ & $2.7_{-0.4}^{+0.9}$ & 574.6/476  \\ 
\vspace{.001in} \\
3.50--5.00 & $23_{-1}^{+2}$ & $0.48_{-0.01}^{+0.01}$ & $353_{-18}^{+18}$ & $1.3_{-0.2}^{+0.6}$ & 439.8/476  \\ 
\vspace{.001in} \\
5.00--7.00 & $22_{-2}^{+2}$ & $0.47_{-0.02}^{+0.02}$ & $298_{-19}^{+20}$ & $1.3_{-0.2}^{+0.7}$ & 513.6/476  \\
\enddata
\tablenotetext{a}{Surface brightness $I$ in units of $10^{-9}$ photons sec${}^{-1}$ cm${}^{-2}$ arcsec${}^{-2}$}
\end{deluxetable}

\clearpage

\begin{deluxetable}{crrrrrrrr}
\tablecolumns{9}
\tablewidth{0pc}
\tablecaption{Double Beta-Model Fits\label{Table-3}}
\tablehead{ 
\colhead{E [keV]} & \colhead{$R_{\rm c1}$} [\arcsec] &
\colhead{$\beta_1$}  & \colhead{$I_{\rm 01}$\tablenotemark{a}} &
\colhead{$R_{\rm c2}$} [\arcsec] & \colhead{$\beta_2$}  &
\colhead{$I_{\rm 02}$\tablenotemark{a}} & \colhead{$I_{\rm
B}$\tablenotemark{a}}  & \colhead{$\chi^2/\rm{DOF}$}  }

\startdata 
0.29--7.00 & $1.5_{-0.2}^{+0.7}$ & $0.41_{-0.05}^{+0.13}$ & $3819_{-637}^{+603}$ & $20_{-1}^{+1}$ & $0.49_{-0.01}^{+0.02}$ & $3367_{-225}^{+258}$ & $14_{-2}^{+2}$ & 553.7/473  \\ 
\vspace{.001in} \\
0.29--2.00 & $2.3_{-0.4}^{+1.3}$ & $0.45_{-0.07}^{+0.09}$  & $2227_{-304}^{+217}$ & $17_{-1}^{+2}$ & $0.49_{-0.01}^{+0.02}$ & $1907_{-318}^{+864}$ & $10_{-2}^{+2}$ & 536.1/473  \\ 
\vspace{.001in} \\
2.00--7.00  & $1.1_{-0.2}^{+0.2}$ & $0.41_{-0.05}^{+0.15}$ & $1583_{-158}^{+91}$ & $21_{-1}^{+2}$ & $0.49_{-0.01}^{+0.02}$  & $1959_{-326}^{+933}$ & $7_{-1}^{+1}$ & 535.3/473  \\  

\enddata
\tablenotetext{a}{Surface brightness $I$ in units of $10^{-9}$ photons sec${}^{-
1}$ cm${}^{-2}$ arcsec${}^{-2}$}
\end{deluxetable}

\clearpage

\begin{deluxetable}{crrrrr}
\tablecolumns{6}
\tablewidth{0pc}
\tablecaption{Mass Estimates for R $<$ 0.5 Mpc\label{Table-4}}
\tablehead{                
\colhead{Model}           &
\colhead{$\rm{kT}$}            &
\colhead{$\rm{n}_0$}           &
\colhead{$\rm{M}_{\rm{gas}}$}       &
\colhead{$\rm{M}_{\rm{tot}}$}       &
\colhead{$\rm{M}_{\rm{arc}}$}       \\
\colhead{}                &
\colhead{[keV]}           &
\colhead{[cm$^{-3}$]}       &
\colhead{[10$^{13} \rm{M}_\odot$]}       &
\colhead{[10$^{13} \rm{M}_\odot$]}       &
\colhead{[10$^{13} \rm{M}_\odot$]}       }
\startdata
Single T &       $7.6_{-0.3}^{+0.3}$  & $ 0.070_{-0.001}^{+ 0.001}$ & $  5.6_{ -0.4}^{+  0.5}$ & $ 18.9_{ -0.9}^{+  1.0}$ & $  1.9_{ -0.1}^{+  0.1}$ \\
      2T &      $10.0_{-0.9}^{+0.8}$  & $ 0.069_{-0.001}^{+ 0.001}$ & $  5.5_{ -0.4}^{+  0.4}$ & $ 24.8_{ -2.1}^{+  2.1}$ & $  2.5_{ -0.2}^{+  0.2}$ \\
   1T+CF &   $12.08_{-0.78}^{+0.01}$  & $ 0.069_{-0.001}^{+ 0.001}$ & $  5.5_{ -0.4}^{+  0.4}$ & $ 29.1_{ -1.3}^{+  1.4}$ & $  2.9_{ -0.1}^{+  0.1}$ \\
\enddata
\end{deluxetable}
\clearpage

\begin{figure}
\begin{center}
\includegraphics[height=6.5in,angle=270]{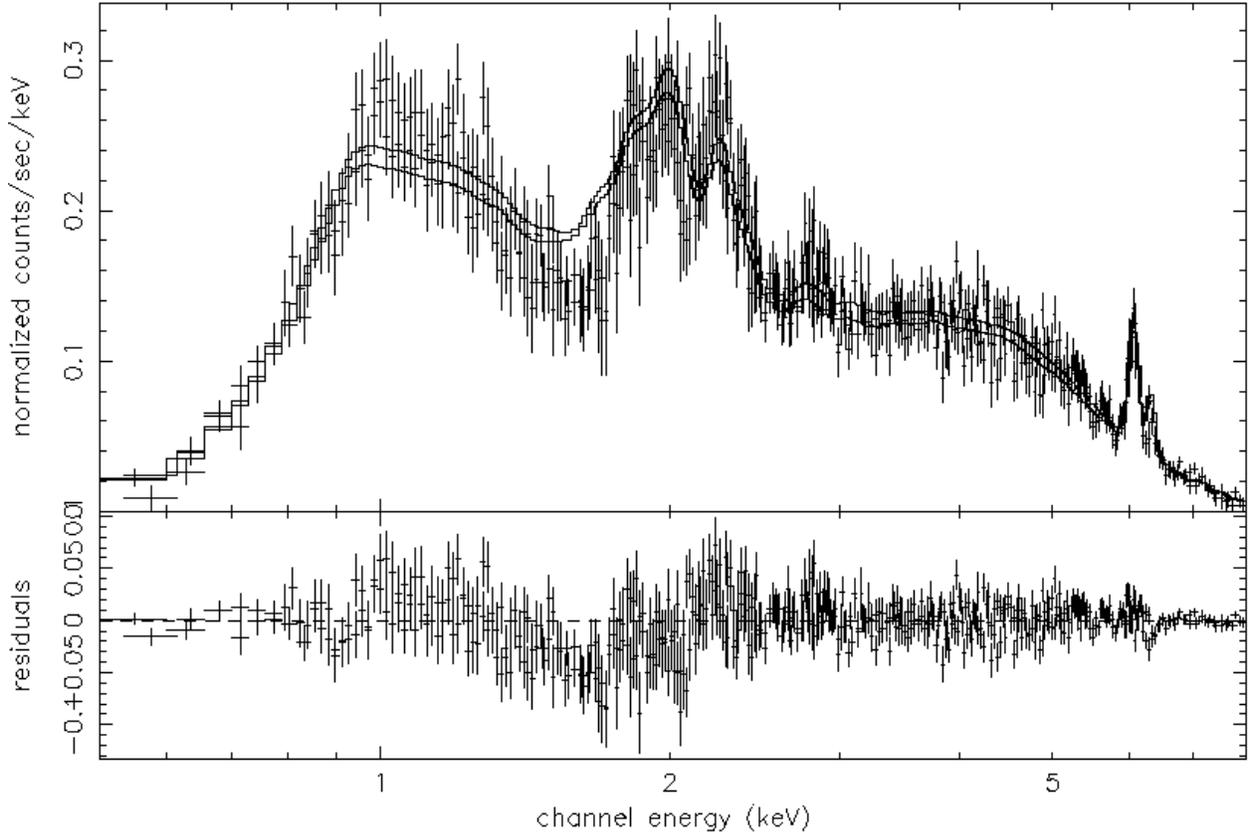}
\figcaption{{\bf Core Spectrum with Model Comparisons.}  XSPEC plots
showing a spectrum fit with a two temperature model.  This fit
resulted in an ambient cluster temperature of ${8.7}^{+0.5}_{-0.5}$
keV, a low temperature component of ${1.2}^{+0.1}_{-0.1}$ keV, an
abundance of ${0.52}^{+0.05}_{-0.05}$ solar,  and a reduced $\chi^2$
of 1.02 for 929 degrees of freedom.  The spectrum was extracted from
a circle centered on the emission peak with a radius of $2\arcmin$
($\sim$300 kpc), and was grouped to include 100 counts per bin for
clarity. \label{Figure-1}}    
\end{center}
\end{figure}

\clearpage

\begin{figure}

\includegraphics*[bb=30 165 522 610, height=5.25in]{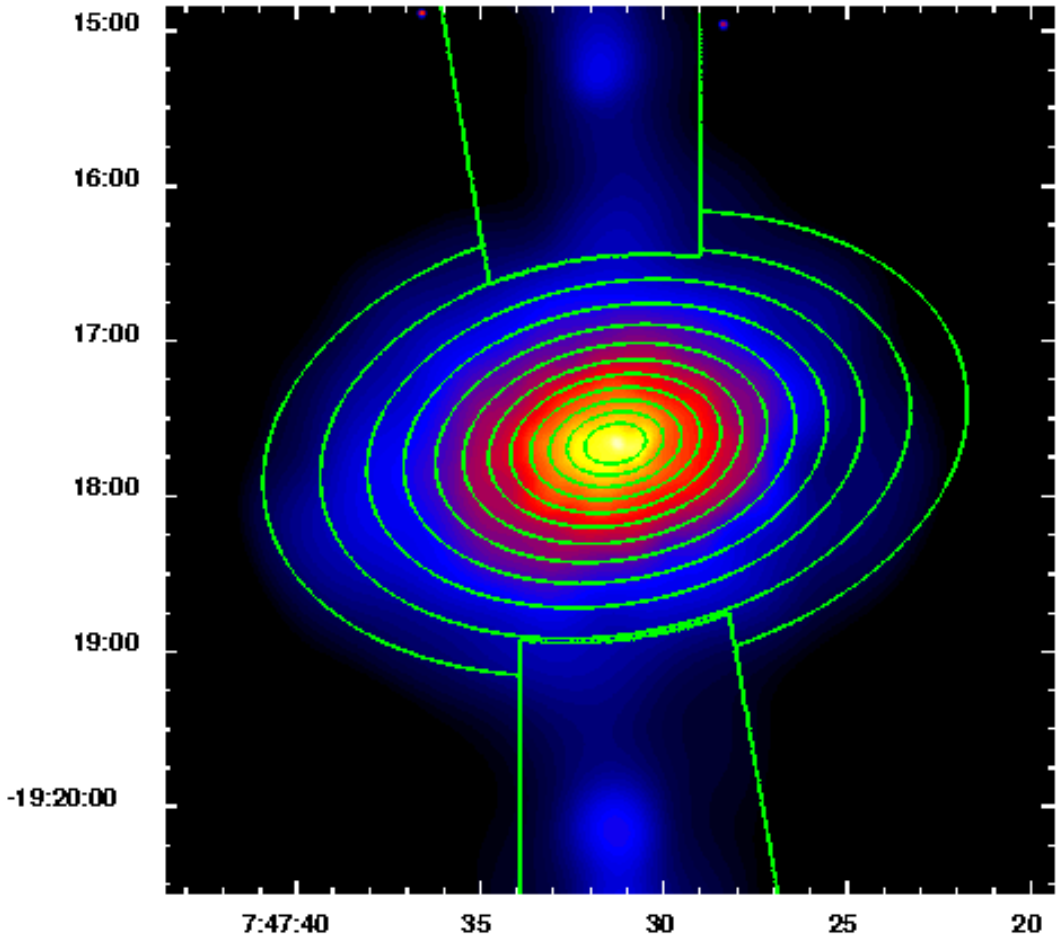}
\figcaption{{\bf Core Flux Image.}  Adaptively smoothed image of the
core of PKS 0745-191, made from the longer of the two observations.
Contours highlight elliptical regions from which spectra were
extracted to produce temperature and abundance profiles, along with
regions excluded from our spectral analysis due to possible
diffracted photon contamination.  The peak of emission lies within
$0.5\arcsec$ (the spatial resolution of Chandra) of the cD
galaxy.\label{Figure-2}}  
\end{figure}

\clearpage
\begin{figure}

\plotone{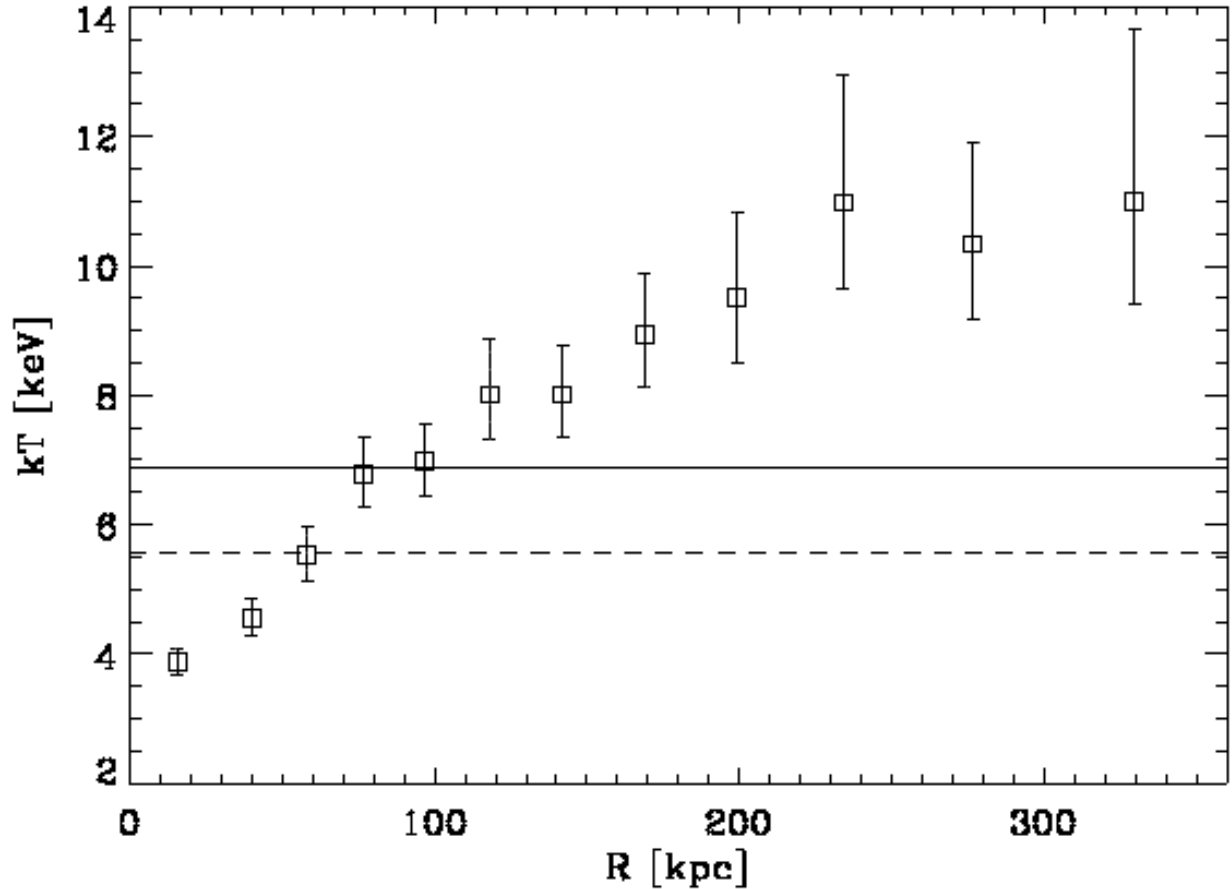}
\figcaption{{\bf Single Temperature Profile.}  Radial temperature
profile determined by fitting, in XSPEC, single temperature spectral
models to spectra extracted from elliptical annuli selected to include
a minimum of 3000 counts in the 0.29-7.0 keV band (see Figure 1).
Error bars represent 90\% confidence limits.  The solid line
represents the temperature obtained by fitting $0-2\arcmin$
Chandra data with a single temperature model.  The dotted line
indicates the temperature reported by \citet{allen-PKS} as a result of
their $0-2\arcmin$ ($\sim$310 kpc) single temperature ASCA analysis of
PKS 0745-191.     
\label{Figure-3}}    
\end{figure}

\clearpage

\begin{figure}

\plotone{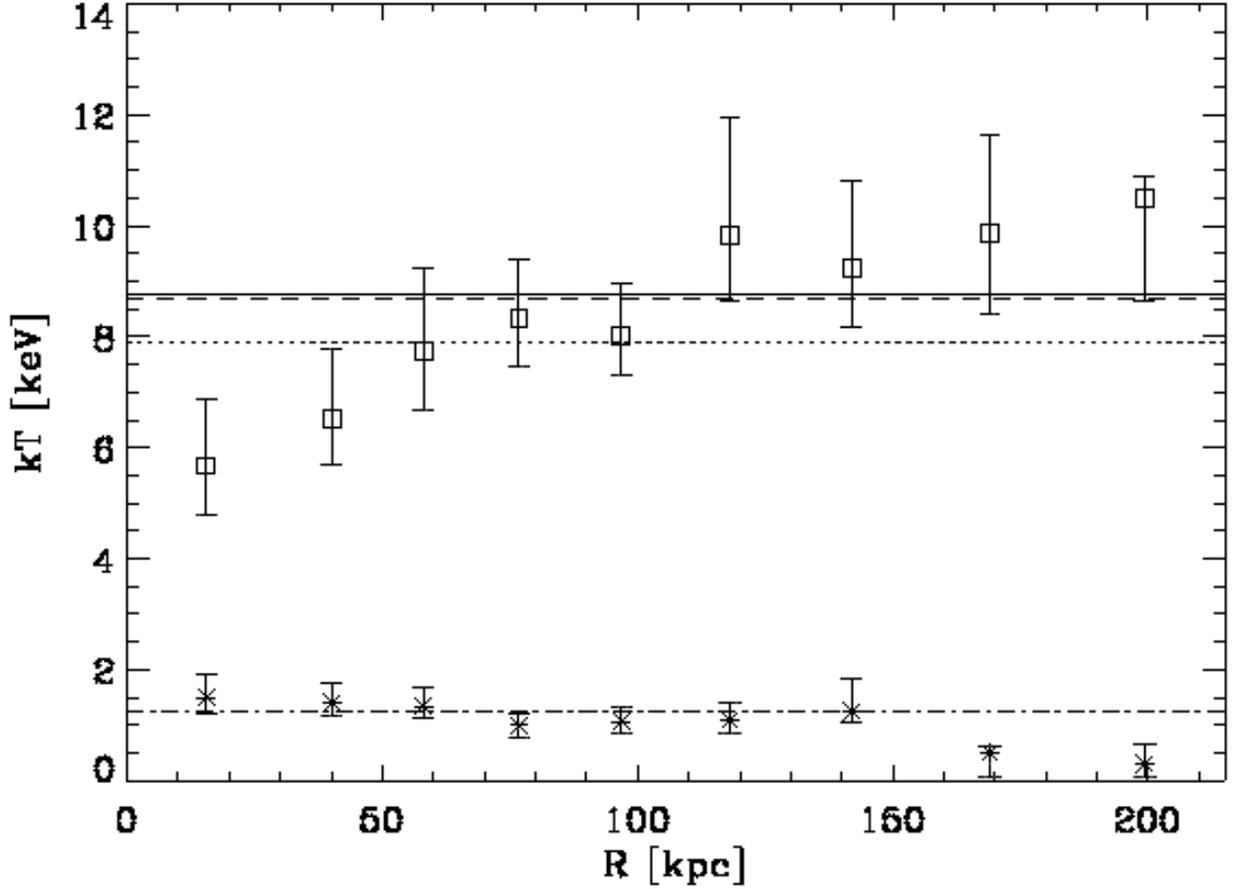}
\figcaption{{\bf Two Temperature Profile.}  Radial temperature profile
determined by fitting a two temperature spectral model to spectra
extracted from the elliptical annuli discussed in {Figure-3}.  Error
bars represent 90\% confidence limits.  The solid line indicates the
best fitting (high) temperature from a $0-2\arcmin$ Chandra two
temperature analysis of this cluster, and the dot-dash line
illustrates the best fitting low temperature component of the same
analysis.  The dashed line represents the value quoted from the
BeppoSAX cooling flow analysis of the inner $2\arcmin$ of PKS
0745-191, and the dotted line indicates the ASCA SIS0 cooling flow
model result for the same region.  The low temperature component of
the two temperature models could not be constrained at radii larger
than 215 kpc ($1.4\arcmin$).     
\label{Figure-4}}  
\end{figure}

\clearpage

\begin{figure}

\plotone{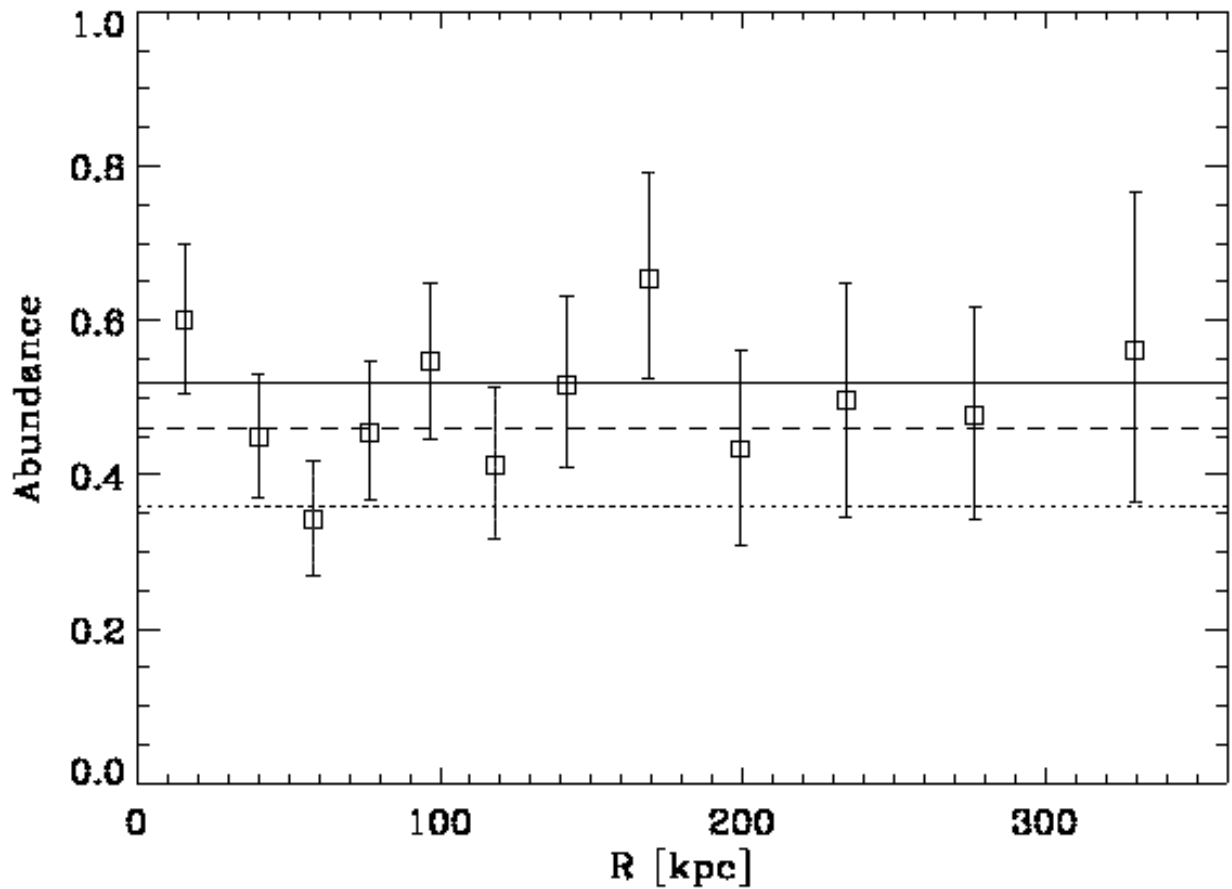}
\figcaption{{\bf Abundance Profile.}  Abundance profile determined
with a single phase spectral model.  The abundances that we
present here are consistent (within their errors) with the
corresponding abundances resulting from our two temperature analysis.
Although not as well constrained as the temperature profile, our
analysis is consistent with a constant abundance of roughly 0.5 solar 
throughout the inner region of PKS 0745-191.  The solid line is the
abundance value resulting from an $0-2\arcmin$ two temperature
analysis, the dashed line is the BeppoSAX $0-2\arcmin$ cooling flow
model result, and the dotted line represents the outcome of the ASCA
multiphase analysis of that same region.\label{Figure-5}}      
\end{figure}

\clearpage

\begin{figure}

\includegraphics[width=5in, angle=270]{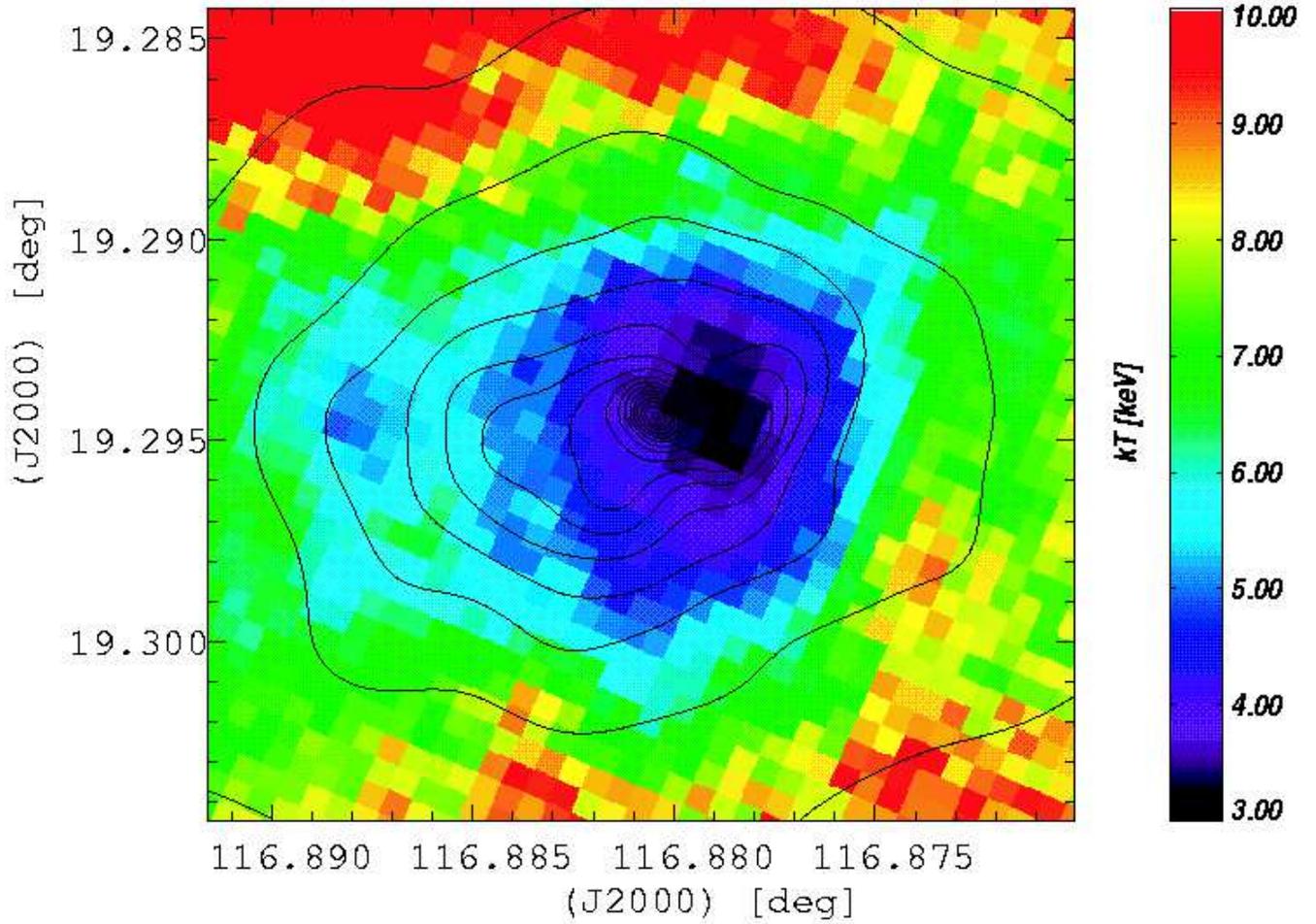}
\figcaption{{\bf Temperature Map.} 
The temperature map for the central 100$^{''}$x100$^{''}$ region
of PKS0745-191 (OBSID 510) centered on the peak of cluster emission. 
The contours show the adaptively smoothed X--ray surface brightness.
The color bar indicates the temperature scale in keV.
The statistical error in the map is $\sim$0.8 keV at the 90\%
confidence level.
\label{Figure-6}}       
\end{figure}
\clearpage

\begin{figure}

  \includegraphics[width=5in, angle=270]
{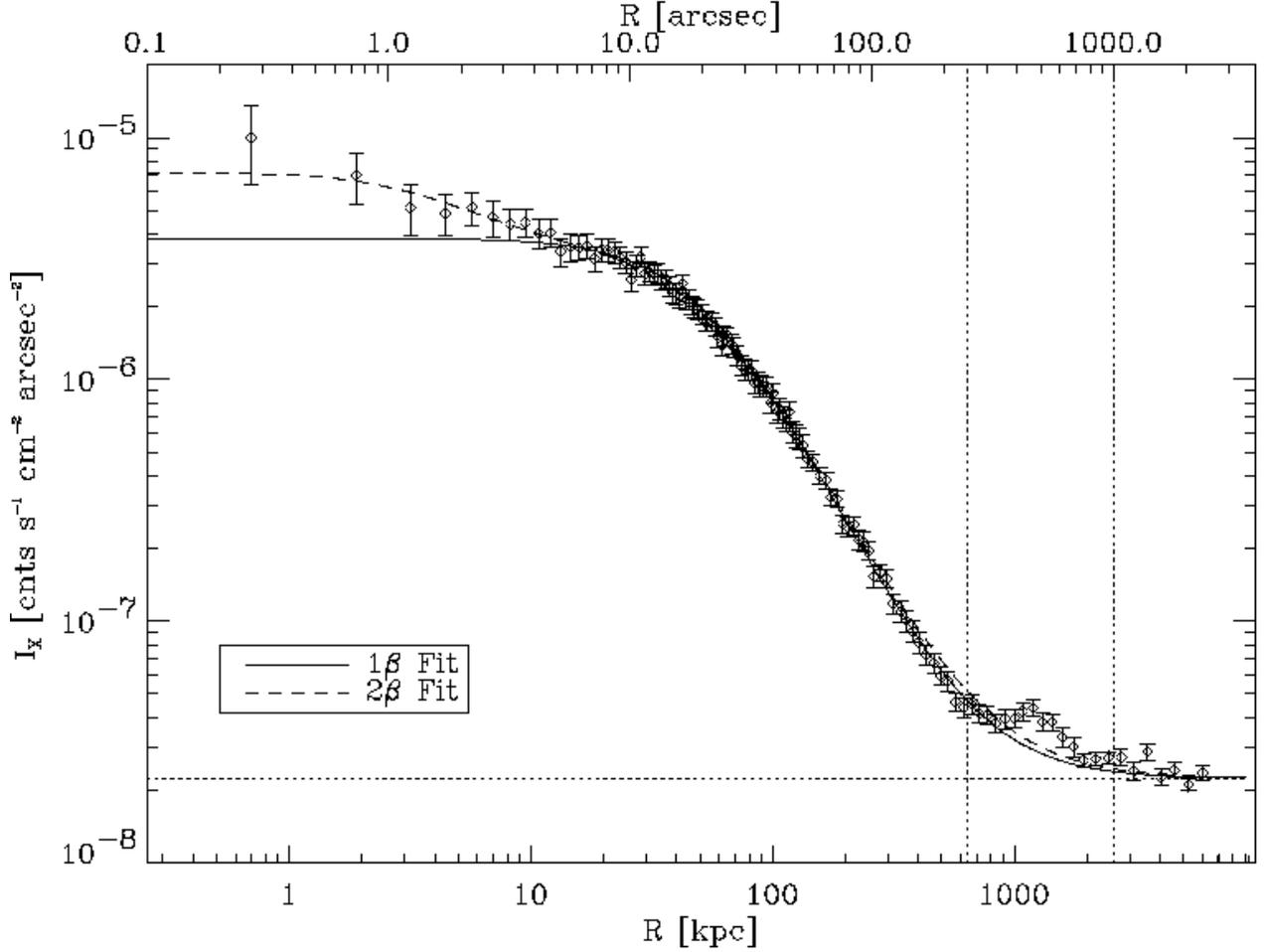}
\figcaption{{\bf Surface Brightness Profile.} Radial surface
brightness for the 0.29-7.0 keV band accumulated in $1\arcsec$ annular
bins.  A solid line traces the best single $\beta$ model fit and
the dashed line illustrates the improved fit obtained by fitting this
data with a double $\beta$ model.  The horizontal dotted line
represents the best fitting background value.  Two vertical dotted
lines indicate a region of data which was not included in the fit, due
to residual contamination from first order grating photons.  The
remaining data was well fit by a double $\beta$ model with an inner
core radius of 1.5 $\arcsec$ (3.86 kpc) and beta value of 0.41, and
an outer core radius of 20 $\arcsec$ (51.42 kpc) and $\beta$ of
0.49\label{Figure-7}}       
\end{figure}
\clearpage

\begin{figure}

\plotone{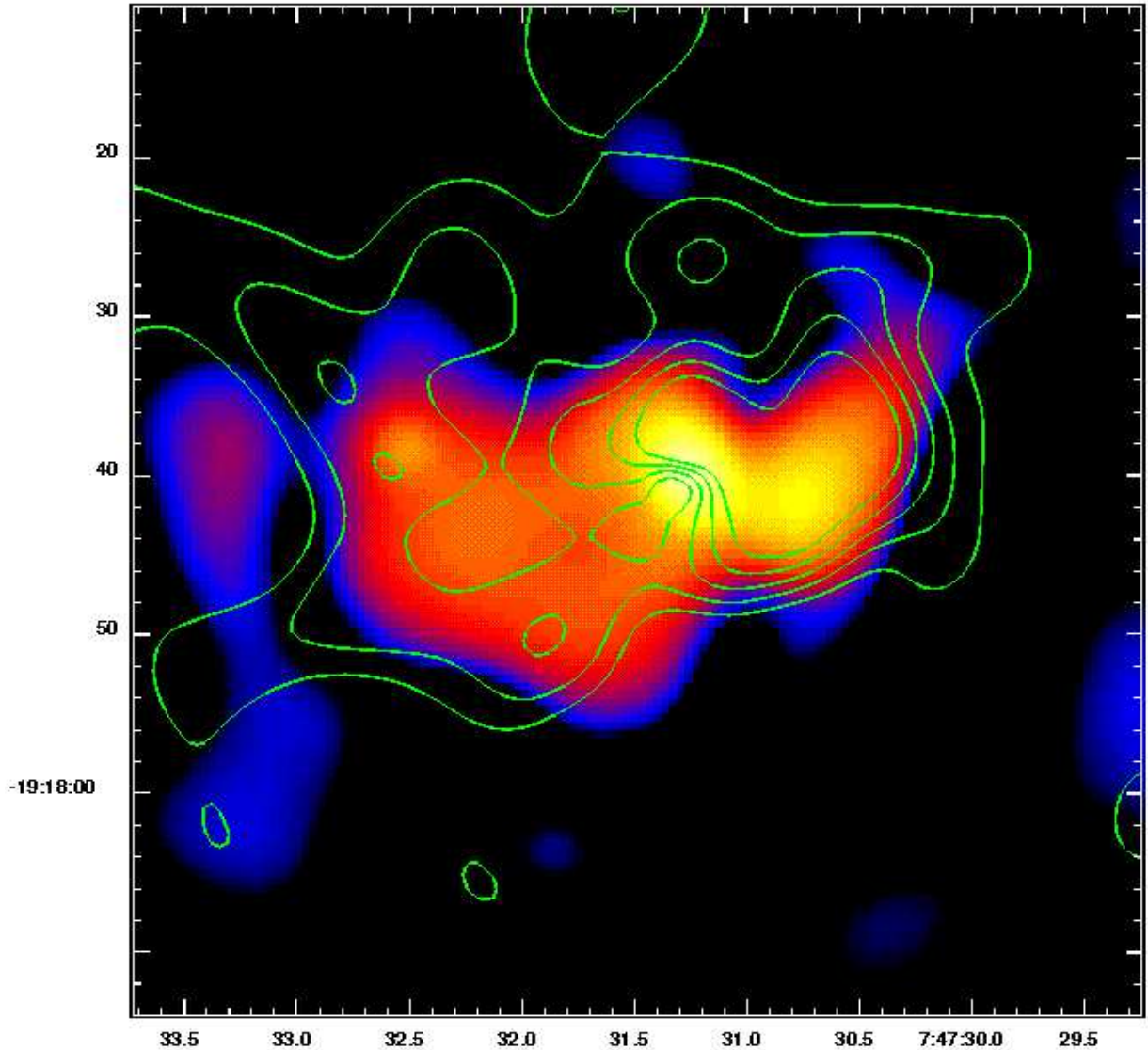}
\figcaption{{\bf Core Residuals.}  Image of PKS 0745-191 indicating
core substructure.  A highly smoothed image of this region was
subtracted from an adaptively smoothed image of PKS 0745-191
to enhance X-ray substructure in the core.  The peak of emission lies
at the location of the cD galaxy, and has a surface brightness which
is 65\% of the adaptively smoothed surface brightness at that point.
Five contours of constant hardness ratio (\S \ref{ss:hardness}) are
spaced linearly over the range 0.06 to 0.18, and illustrate variations
in the value $\frac{\rm{I_{s}-I_{h}}}{\rm{I_{tot}}}$.\label{Figure-8}}   
\end{figure}

\clearpage

\begin{figure}

\begin{center}
\includegraphics*[bb=-27 141 640 653, height=5.0in]{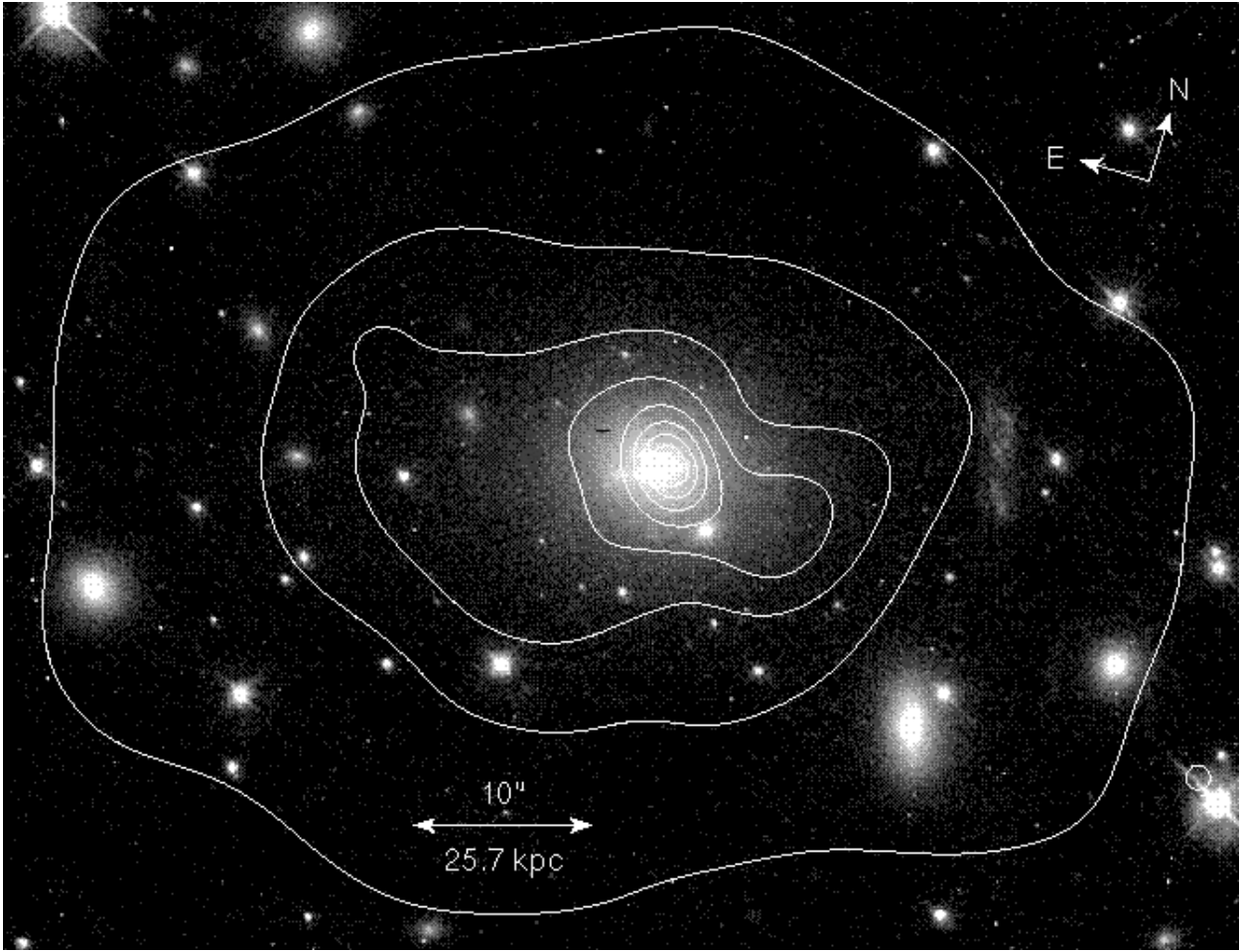}
\figcaption{{\bf Optical Overlay.} HST image of PKS 0745-191, spanning
  the wavelength range 4789.0 $\rm {\AA}$ to 6025.0 $\rm {\AA}$.  Chandra
  X-ray isophotes (see \S\ref{ss:imag-substr} for details) from an
  adaptively smoothed core image, composed of data from obsid 510 in
  the 0.29-7.0 keV range, are overlaid in white.  The X-ray peak is
  within $0.5\arcsec$ of the optical center of the cD galaxy.  Note
  the presence of the gravitational arc directly to the right of the
  cD.\label{Figure-9}} 
\end{center}   
\end{figure}

\clearpage

\begin{figure}

\plotone{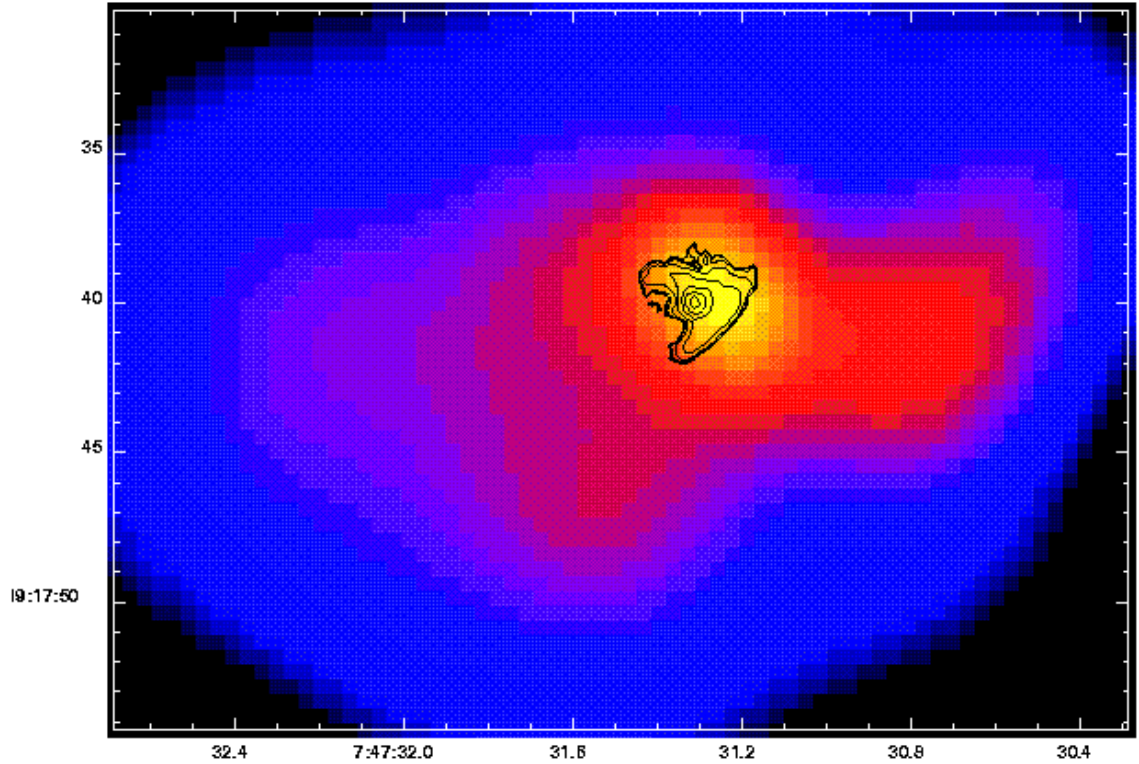}
\figcaption{{\bf Radio Overlay.} Adaptively smoothed X-ray image,
composed of data from obsid 510 in the 0.29-7.0 keV range.  VLA radio
contours (\citet{radio}) (see \S \ref{ss:imag-substr} for details) of PKS 0745-191
are overlaid in black.  The peak of X-ray emission lies with
$0.5\arcsec$ of the central radio source.\label{Figure-10}}  
\end{figure}



\clearpage



\begin{thebibliography}{}

\bibitem[Allen(1998)]{allen} Allen, S. W. 1998, MNRAS, 296, 392.

\bibitem[Allen, Fabian \& Kneib(1996)]{allen-PKS} Allen, S. W.,
Fabian, A. C. \& Kneib, J. P. 1996, MNRAS, 279, 615. 

\bibitem[Anninos \& Norman (1996)]{an96}Anninos, P., \& Norman,
M.L., 1996, ApJ, 459, 12.

\bibitem[Arnaud(1996)]{xspec} Arnaud, K. A. 1996, ADASS, 101, 5.

\bibitem[Arnaud et al.(1987)]{ar} Arnaud, K. A. et al. 1987, MNRAS, 227,
241. 

\bibitem[De Grandi \& Molendi(1999)]{DM99} De Grandi, S. \& Molendi,
S. 1999, A\&A, 351, L45. 

\bibitem[Dickey \& Lockman(1990)]{nh} Dickey, J. M. \& Lockman
F. J. 1990, ARA\&A, 28, 215.

\bibitem[Eke, Cole \& Frenk (1996)]{ecf96}Eke, V.R., Cole, S., \&
Frenk, C.S., 1996, MNRAS, 282, 263.

\bibitem[Eke, Navarro \& Frenk (1998)]{enf98}Eke, V.R., Navarro,
J., \& Frenk, C.S., 1998, preprint.

\bibitem[Evrard(1990)]{ev90} Evrard, A., 1990, ApJ, 363, 349.

\bibitem[Evrard et al.(1996)]{emn96} Evrard, A.E., Metzler, C.A., \&
Navarro, J., F. 1996, ApJ, 469, 494. 

\bibitem[Fabian et al.(1985)]{fabian} Fabian, A. C. et al. 1985, MNRAS,
216, 923.

\bibitem[Fabian (1994)]{fab94} Fabian, A.C., 1994, \araa, 32, 277.

\bibitem[Frenk et al.(1990)]{frenk90}Frenk, C.S. et al. 1990, ApJ, 351, 10.

\bibitem[Houck \& DeNicola(2000)]{houck} Houck, J.C., \& DeNicola, L.A. 2000, ASP Conf. Ser., 216, 591. 

\bibitem[Houck et al.(2001)]{tmap} Houck, J.C., Davis, D. S., \& Wise,
M. W. 2002, ApJ, in preparation. 

\bibitem[Ikebe et al.(1999)]{cent} Ikebe, Yasushi et al. 1999, ApJ,
525, 58. 

\bibitem[Jones \& Forman(1984)]{beta} Jones, C. \& Forman, W. 1984,
ApJ, 276, 38.

\bibitem[Kaiser \& Squires (1993)]{ks93}Kaiser, N., \& Squires,
G., 1993, ApJ, 404, 441.

\bibitem[Lynds \& Petrosian(1988)]{lynds} Lynds, R. \& Petrosian,
V. 1988, Large Scale Structures of the Universe: Proc. of the 130th
Symp. of the IAU, 130, 467.

\bibitem[Mathiesen \& Evrard (2001)]{me01}Mathiesen, B.F., \&
Evrard, A.E., 2001, ApJ, 546, 100.

\bibitem[McNamara et al.(2000)]{mcnamara} McNamara, B. R. et al. 2000,
ApJ, 534, L135. 

\bibitem[Navarro, Frenk \& White (1995)]{nfw95}Navarro, J., Frenk,
C.S., \& White, S.D.M., 1995, MNRAS, 275, 720.

\bibitem[Peterson et al.(2000)]{xmm} Peterson, J. R. et al. 2000,
preprint (astro-ph/0010658).

\bibitem[Peterson et al.(2001)]{peterson} Peterson, J. R. et al. 2001,
A\& A, 365, L104.

\bibitem[Richstone, Loeb \& Turner (1992)]{rlt92}Richstone, D.,
Loeb, A., \& Turner, E.L., 1992, ApJ, 393, 477.

\bibitem[Sarazin (1988)]{cls88}Sarazin, C.L., 1988, ``X-ray
Emission from Clusters of Galaxies'', Cambridge University Press.

\bibitem[Schindler (1996)]{sch96}Schindler, S., 1996, A\& A, 305,
756.

\bibitem[Soucail et al.(1988)]{soucail} Soucail, G. et al. 1988, A\&A,
191, L19. 

\bibitem[Taylor, Barton, \& Ge(1994)]{radio} Taylor, G. B., Barton, E.
J. \& Ge, J. 1994, AJ, 107, 1942.

\bibitem[Tyson, Valdes \& Wenk 1990]{tvw90}Tyson, J., Valdes, F.,
\& Wenk, R., 1990, ApJ, 349, L19.

\bibitem[Wise et al.(2002)]{wise} Wise, M. W. et al. 2002, in
  preparation. 

\bibitem[Xue \& Wu(2000)]{doubeta} Xue, Yan-Jie \& Wu, Xiang-Ping
2000, MNRAS, 318, 715. 

\end{thebibliography}
\end{document}